\documentclass{PoS}
\title{Magnetic field measurements at milliarcsecond resolution around massive young stellar objects.}

\ShortTitle{Magnetic field measurements at mas resolution around massive YSOs}

\author{\speaker{Gabriele SURCIS}\\
       Joint Institute for VLBI in Europe\\
       E-mail: \email{surcis@jive.nl}}

\author{Wouter H.T. VLEMMINGS\\
        Chalmers University of Technology, Onsala Space Observatory\\
        E-mail: \email{wouter.vlemmings@chalmers.se}}

\author{Huib J. VAN LANGEVELDE\\
        Joint Institute for VLBI in Europe and Sterrewacht Leiden, Leiden University\\
        E-mail: \email{langevelde@jive.nl}}

\author{Busaba HUTAWARAKORN KRAMER\\
        Max-Planck Institut f\"{u}r Radioastronomie\\
        E-mail: \email{bkramer@mpifr-bonn.mpg.de}}

\author{Anna BARTKIEWICZ\\
        Centre\,for\,Astronomy,\,Faculty\,of\,Physics,\,Astronomy\,and\,Informatics,\,Nicolaus\,Copernicus\,University\\
        E-mail: \email{annan@astro.uni.torun.pl}}

\author{Hans ENGELKAMP\\
        High\,Field\,Magnet\,Laboratory,\,Institute\,for\,Molecules\,and\,Materials,\,Radboud\,University\,Nijmegen\\
        E-mail: \email{h.engelkamp@science.ru.nl}}

\def\meth {CH$_{3}$OH}
\abstract{Magnetic fields have only recently been included in theoretical simulations of
high-mass star formation. The simulations show that magnetic fields can play a
crucial role not only in the formation and dynamics of molecular outflows, but
also in the evolution of circumstellar disks. Therefore, new 
measurements of magnetic fields at milliarcsecond resolution close to massive 
young stellar objects (YSOs) are fundamental for providing new input for numerical
simulations and for understanding the formation process of massive stars.
The polarized emission of 6.7 GHz \meth ~masers allows us to investigate the 
magnetic field close to the massive YSO where the outflows and disks
are formed. Recently, we have detected with the EVN \meth ~maser
polarized emission towards 10 massive YSOs. From a first
statistical analysis we have found evidence that magnetic fields
are primarily oriented along the molecular outflows.
To improve our statistics we are carrying on a large observational EVN campaign for a total of
19 sources, the preliminary results of the first seven sources are presented in this contribution. 
Furthermore, we also describe our efforts to estimate the Land\'{e} 
$g$-factors of the \meth ~maser transition to determine the 
magnetic field strength from our Zeeman-splitting measurements.}
\FullConference{12th European VLBI Network Symposium and Users Meeting\\
		7-10 October 2014\\
		Cagliari, Italy}
\begin{document}

\def\hii {H\,{\sc ii}}
\def\water {H$_2$O}
\def\meth {CH$_{3}$OH}
\def\kms{km\,s$^{-1}$}
\def\solmass {\hbox{M$_{\odot}$}}
\def\d {$^{\circ}$}
\def\mjyb{mJy\,beam$^{-1}$}
\def\dvz {$\Delta V_{\rm{Z}}$}

\section{Introduction}
The whole process that governs low-mass star formation ($\rm{M}<8$~\solmass) is nowadays thought
to be fairly well understood. The low-mass stars are formed through gravitational collapse of bound cores 
that are created by the fragmentation of molecular clouds. During the formation of the star an accretion disk 
around the central protostar is formed and a jet/bipolar outflow is launched perpendicular to the disk.
Furthermore, the magnetic field is thought to play an important role in slowing 
the collapse, in transferring the angular momentum, and in powering the outflow (e.g., \cite{mck07}). However,
there still exists an open debate on whether the orientation of the magnetic field aligns with respect
to the orientation of molecular outflow. Recently, two independent surveys of dust polarized emission towards 
low-mass protostellar cores showed opposing results. One found no correlation between magnetic field 
orientation and outflow axis in low-mass young stellar objects (YSOs) \cite{hul13}, while the other one
found a good alignment \cite{cha13}.\\
\indent The \textit{core accretion} model describes the formation of high-mass stars ($\rm{M}>8$~\solmass) as 
a scaled-up version of the low-mass star formation (e.g., \cite{mck03}). Recent theoretical simulations
have suggested that magnetic fields might play a role in massive star formation as important as in low-mass
star formation. Indeed, the simulations begin to reproduce the observations only when the 
magnetic field is taken into consideration (e.g., \cite{pet11}, \cite{sei12}, \cite{sei15}). Similarly to
the low-mass star formation case, conflicting results on the orientation of the magnetic field with respect to
the outflow orientation have also been found. Based on the observations of dust polarized emission towards 
a sample of 21 sources, no correlation between outflow axis and magnetic field orientation has been found at
arcsecond resolution 
($\sim10^3$~au) in massive star-forming regions (SFRs) \cite{zha14}. On the contrary, by observing the 
polarized emission of 6.7 GHz \meth ~masers with the European VLBI Network\footnote{The European VLBI 
Network is a joint facility of European, Chinese, South African and other radio astronomy institutes funded 
by their national research councils.} we have found evidence that on scales of 10-100 au magnetic field
around massive YSOs is preferentially oriented along the outflow \cite{sur13}. This is supported by a 
Kolmogorov-Smirnov (K-S) test performed on a sample of nine sources that shows a probability of 10\% that the 
distribution of the projected angles $|\rm{PA}_{\rm{outflow}}-\langle\Phi_{\rm{B}}\rangle|$ is drawn from a
random distribution (see Fig.~\ref{cdf}). Here, $\rm{PA}_{\rm{outflow}}$ is the orientation of the outflow
and $\langle\Phi_{\rm{B}}\rangle$ is the error weighted orientation of the magnetic field~\cite{sur13}. \\
\indent Furthermore, providing new measurements of magnetic fields strength at mas resolution close to the 
massive YSOs by measuring the Zeeman splitting of the 6.7 GHz \meth ~maser emission is fundamental to 
verify and/or improve the numerical simulations of massive star formation. Even though in the last years 
Zeeman-splitting measurements of \meth ~maser emission have been made, the exact proportionality between
the measured splitting and the magnetic field is still uncertain (\cite{vle11}, \cite{sur12}, \cite{sur13}).
Therefore, it is of great importance to measure the still unknown Land\'{e} g-factors for the \meth ~molecule.
\section{The flux-limited sample}
\begin{figure*}[ht]
\centering
\includegraphics[width = 7.5 cm]{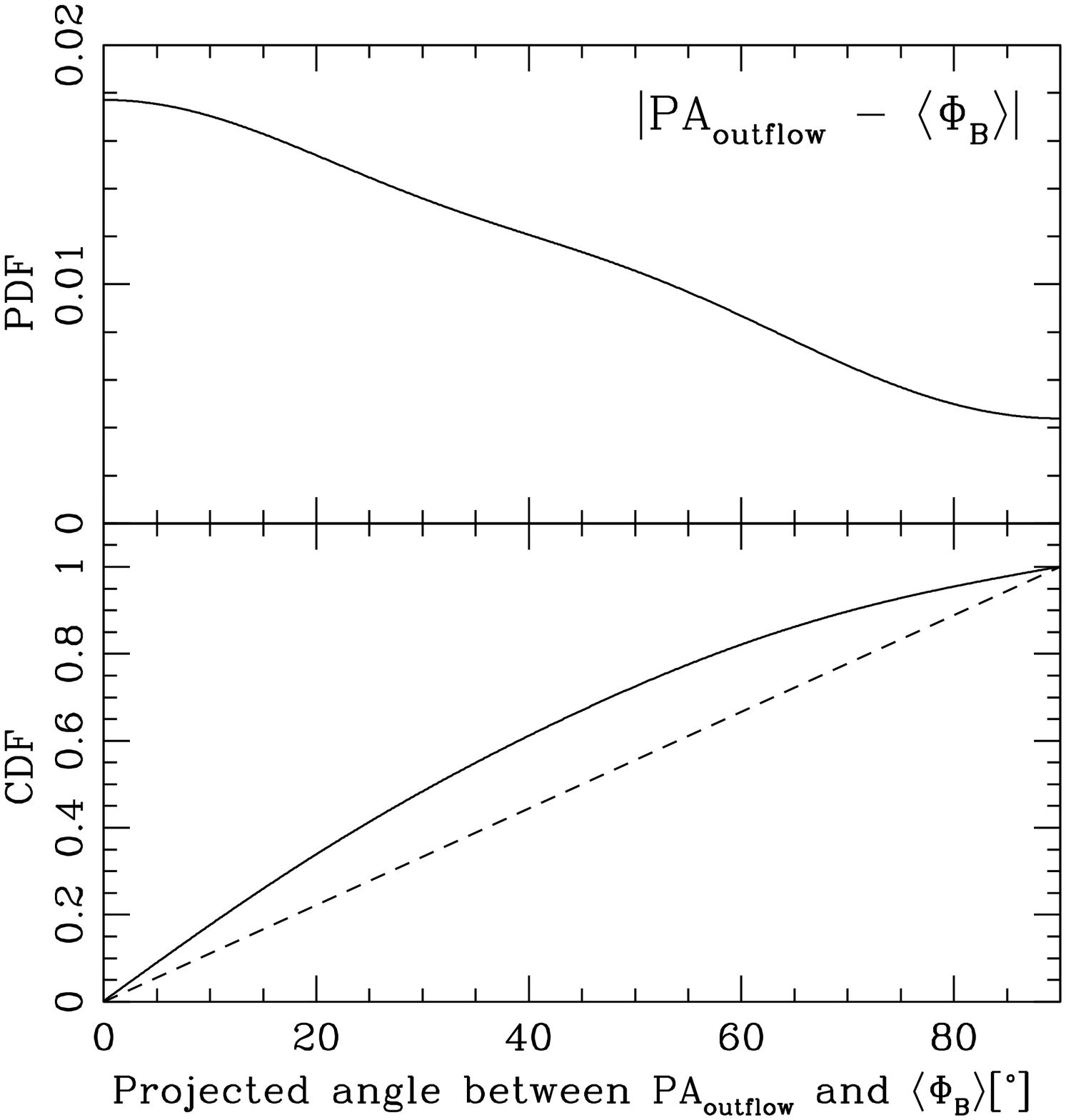}
\includegraphics[width = 7.5 cm]{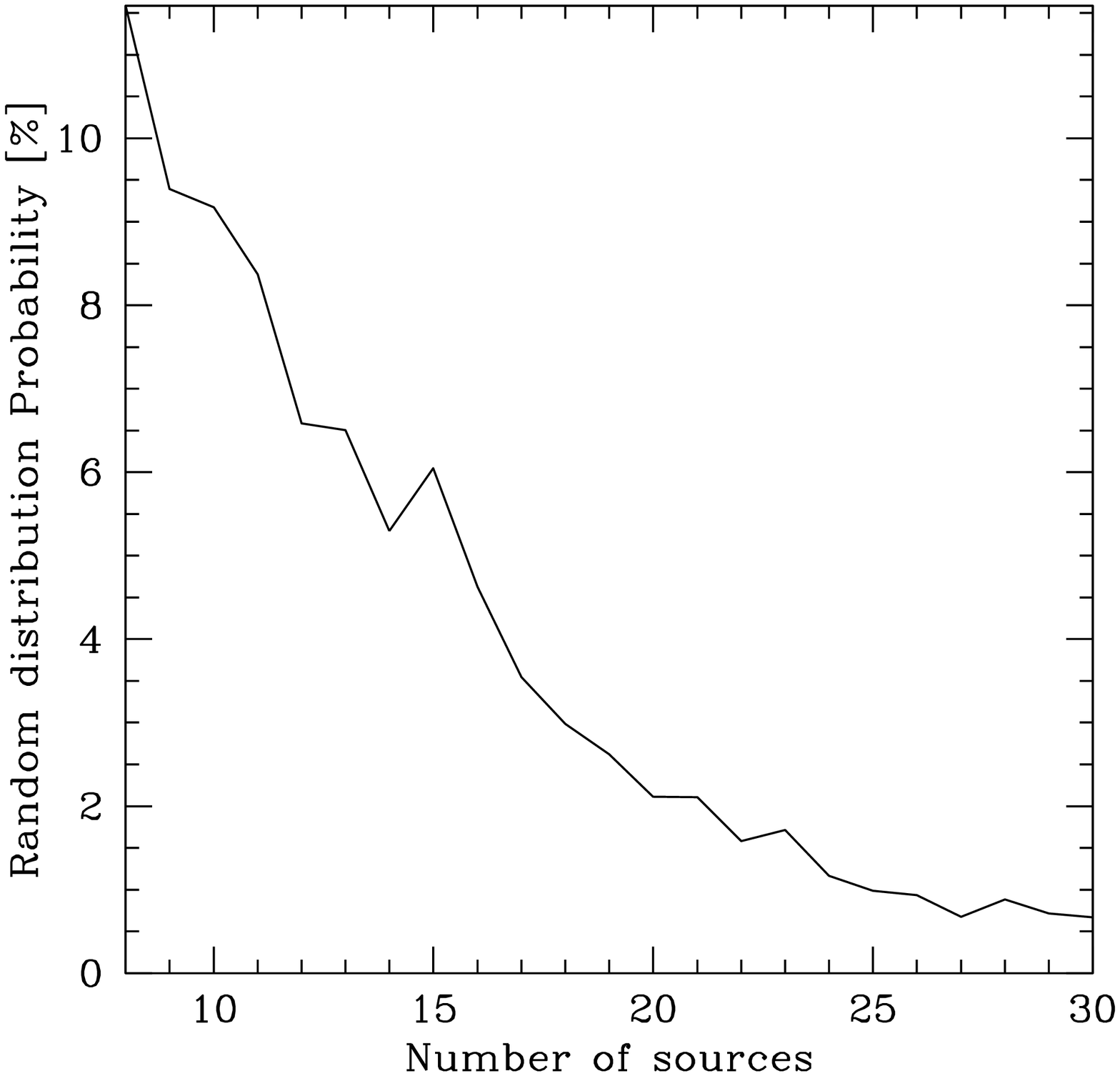}
\caption{\footnotesize{\textit{Left}. The probability distribution function (PDF, top panel) and the 
cumulative distribution function (CDF, bottom panel) of the projected angle between the outflow axes and 
the magnetic field ($|\rm{PA}_{\rm{outflow}}-\langle\Phi_{\rm{B}}\rangle|$) based on nine sources as 
measured by \cite{sur13}. The dashed line is the CDF for random orientation of outflows and magnetic fields, 
i.e. all angular differences are equally likely. \textit{Right}. The random distribution probability of 
$|\rm{PA}_{\rm{outflow}}-\langle\Phi_{\rm{B}}\rangle|$ determined with a K-S test as function of 
number of sources.}}
\label{cdf}
\end{figure*}
To improve our statistics it is important to enlarge the number of massive SFRs toward which 
$|\rm{PA}_{\rm{outflow}}-\langle\Phi_{\rm{B}}\rangle|$ can be measured; in other words, we need to enlarge the 
number of massive YSOs for which the orientation of the magnetic field at milliarcsecond (mas) resolution is measured.
With a Monte-Carlo simulation we determined the number of sources that we would need to significantly decrease 
the K-S probability, taking into account observational errors in the outflow and magnetic field angles.
Hence we determined the K-S probability between 8 and 30 vectors pairs (or sources). The vectors in a pair represent
$\rm{PA}_{\rm{outflow}}$ and $\langle\Phi_{\rm{B}}\rangle$ and are randomly selected, with the only apriori condition 
that their 3D misalignment is $<30$\d. We found that 
the probability decreases below 1\% for more than 28 sources (Fig~\ref{cdf}).\\
\indent We have selected a flux-limited sample of massive SFRs with declination $>-9$\d ~and a total \meth ~maser 
single-dish flux greater than 50~Jy from the 6.7 GHz \meth ~maser catalogue of \cite{pes05}. To detect circularly 
polarized \meth ~maser emission ($\leq1\%$), we have excluded the regions hosting \meth ~maser that in recent 
single-dish observations showed a total flux below 20~Jy \cite{vle11}. The total number of massive SFRs of the 
flux-limited sample is thus 31. The polarimetric 6.7 GHz \meth ~maser observations, and the consequent magnetic field 
measurements, of twelve of these SFRs had already been published in recent past (\cite{vle10}, \cite{sur09}, 
\cite{sur11}, \cite{sur12}, \cite{sur13}, \cite{sur14}). Therefore, 19 massive SFRs remain to be observed. We were 
given EVN time to observe all of them at 6.7~GHz. \\
\indent To date 16 out of the 19 sources have been observed in full polarization mode at 6.7~GHz by using eight of 
the EVN antennas (Effelsberg, Jodrell, Onsala, Medicina, Noto, Torun, Westerbork, and Yebes-40\,m), for a total 
observation time of 112~h. The remaining three sources are scheduled to be observed during the second EVN session 
(May--June) of 2015. The results of the first seven observed sources are briefly summarized in Sect.~\ref{res} and 
extensively reported in \cite{sur15}. 
\section{Results}
\label{res}
We have detected a total of 176 \meth ~maser features towards the first seven sources of the flux-limited sample 
(i.e., G24.78+0.08, G25.65+1.05, G29.86-0.04, G35.03+0.35, G37.43+1.51, G174.20-0.08, and G213.70-12.6). We were 
able to determine the orientation of the magnetic field around all the sources but G174.20-0.08, toward which no 
linearly polarized 6.7~GHz \meth ~maser emission was detected at $\geq5\sigma$ (i.e., $\geq 20$~\mjyb). The 
magnetic field is along the outflow (with a misalignment $<30$\d) in three massive YSOs (G25.65+1.05, G35.03+0.35, 
and G213.70-12.6) and is perpendicular to the outflow ($>75$\d) in the remaining massive YSOs (G24.78+0.08, 
G29.86-0.04, and G37.43+1.51).\\
\indent We performed a new statistical analysis by adding to the nine sources previously studied \cite{sur13} the 
new magnetic field measurements made around the sources reported above and around IRAS\,20126+4104 \cite{sur14}. 
Moreover, in our analysis we also add two of the southern sources observed by \cite{dod12} that were recently 
associated with CO-outflows \cite{zha14}. Although the number of sources for which we measure 
$|\rm{PA}_{\rm{outflow}}-\langle\Phi_{\rm{B}}\rangle|$ is twice than that in \cite{sur13} (18 vs. 9), the K-S test 
still shows a probability of 10\%, which is three times larger than expected (3\% from Fig.~\ref{cdf}). If the 
magnetic field aligns with the outflow axis, this probability difference can be due, for instance, to the 
selection criteria chosen to observe the first seven sources, which might not be representative of the whole 
sample. That is, we have by chance observed all the sources that do not show on the plane of the sky an alignment 
of the magnetic field w.r.t. the outflow axis. Nevertheless, because this probability is low our previous 
conclusion can be considered still valid, i.e. the magnetic field close to the central YSO (10-100~au) is 
preferentially oriented along the outflow axis. Of course, given the 10\% probability, the misalignment scenario 
cannot be ruled out. However, before drawing any conclusion, we have to reduce and analyse the data of the 12 remaining
massive SFRs.
\section{Zeeman-splitting coefficient of CH$_{3}$OH maser}
Besides determining the magnetic field orientation around massive YSOs by analyzing the linearly polarized emission of
6.7 GHz \meth ~maser, we are also able to measure the Zeeman splitting of the maser line by detecting the circularly
polarized emission. Considering all the sources observed with the EVN, we measured Zeeman splittings in the range 
$0.4~\rm{ms^{-1}}<\Delta V_{\rm{Z}}<10~\rm{ms^{-1}}$ (\cite{sur09}, 
\cite{sur11}, \cite{sur12}, \cite{sur13}, \cite{sur14}). The magnetic field strength is simply related to \dvz ~by  
$B=\frac{B_{||}}{cos \theta}=\frac{\Delta V_{\rm{Z}}}{\alpha_{\rm{Z}}}$, where $\theta$ is the angle between the magnetic 
field and the maser propagation direction and $\alpha_{\rm{Z}}$ is the Zeeman-splitting coefficient that depends
on the Land\'{e} $g$-factors ($g_{\rm{L}}$) of the maser emission (e.g., \cite{vle11}). By modeling the linearly 
polarized emission we are able to estimate $\theta$ (e.g., \cite{sur11}), but $\alpha_{\rm{Z}}$ still remains uncertain
because of the unknown $g_{\rm{L}}$ of the 6.7 GHz \meth ~maser transition $5_{1}-6_{0}~A^{+}$\footnote{The \meth ~molecule
has two different symmetries E1 and $A^{+}$.} \cite{vle11}. Therefore, 
to provide magnetic field strength values around massive YSOs it is crucial to estimate the $g_{\rm{L}}$ of the 
\meth ~transition $5_{1}-6_{0}~A^{+}$ by measuring in a laboratory the $g_{\rm{L}}$ factors of more accessible \meth 
~molecule transitions or by theoretical calculations. Preferably~both.\\ 
\begin{figure}[t]
\centering
\includegraphics[width = 9.6 cm]{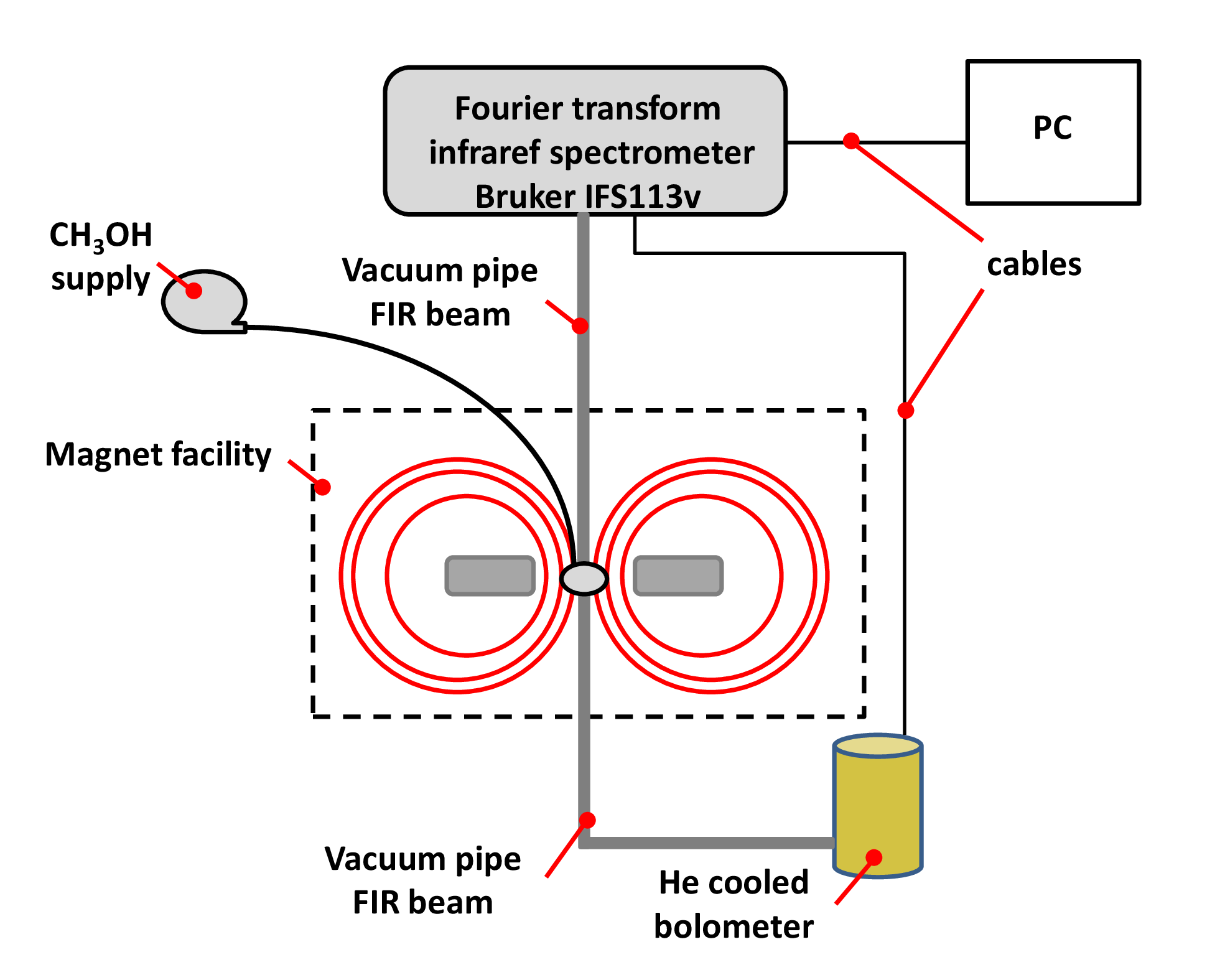}
\caption{\footnotesize{The scheme of the experimental apparatus built at the High Field Magnet Laboratory in Nijmegen
(the Netherlands) to measure the Land\'{e} $g$-factors ($g_{\rm{L}}$) of rotational \meth ~A$^+$ absorption lines in the frequency 
range between 0.6~THz and 1.8~THz. The measured values could be used to deduce the $g_{\rm{L}}$ of the $5_{1}-6_{0}~A^{+}$ 
masing transition.}}
\label{apparatus}
\end{figure}
\indent In November 2013 we set up an experiment by using one of the most powerful magnet facility in Europe, which at that time 
was able to reach a magnetic field of 33.4~T ($3.34\times10^5$~G), at the High Field Magnet Laboratory of the Radboud University\footnote{We acknowledge the support of the HFML-RU/FOM, member 
of the European Magnetic Field Laboratory (EMFL)} (Nijmegen, the Netherlands). We tried to measure the $g_{\rm{L}}$ of many rotational \meth ~$A^+$ absorption lines in the infrared
region (0.6~THz -- 1.8~THz). Because a preliminary study made in 1951 for the $g_{\rm{L}}$ 
of the E1-type \meth ~transitions showed that $g_{\rm{L}}^{\rm{E1}}=0.078+1.88/[J(J+1)]$ \cite{jen51}, 
we suppose that the $g_{\rm{L}}$ will be of the order of 0.1 in the $A^+$ transitions. In order to have a splitting 
of $\sim0.05$~THz, with an experimental resolution at most of $\sim0.001$~THz, magnetic fields up 
to 30~T were required. The scheme of the experimental apparatus that we used is shown in 
Fig.~\ref{apparatus}. A Fourier Transform Infrared Spectrometer generates a far-infrared beam that by traveling into a 
vacuum pipe passes through a sample holder, which is filled with \meth ~gas, that is located at the 
center of the magnet. Here, some of the infrared photons of the beam are absorbed, according to the rotational transitions 
of the molecule, by the \meth ~sample, whose pressure can be varied thanks to a vacuum pump and a \meth ~supply. The 
transmitted infrared beam is then detected by an helium-cooled bolometer and the detected signal is sent to a personal 
computer where the absorption spectrum is displayed. We made several measurements varying both the pressure of the \meth 
~sample (from 10 to 60~mbar) and the magnetic field (from 0 to 30~T). We observed neither the splitting nor the 
broadening of the absorption lines. The none detection of the Zeeman splitting could be due to two main aspects: (1) the splitting
is smaller than the linewidth of the absorption lines, indicating that $g_{\rm{L}}$ for the $A^+$ 
transitions is perhaps smaller than supposed; (2) the quadratic term of the Zeeman effect is not negligible already at small 
fields (few T). Because the measurements are very sensitive to the 
pressure of the \meth ~sample, to circumvent the aforementioned aspects we should design a 
new experiment where we use another source of radiation at lower frequencies or even a laboratory \meth ~maser placed at the 
center of the magnet, which should reach fields on the order of few Tesla. \\
\indent However, there exists another way to estimate the $g_{\rm{L}}$ for the 6.7 GHz \meth ~maser transition apart 
experimentally, that is by theoretically modeling the entire Zeeman effect of the complex CH$_{3}$OH molecule. We contacted the 
Theoretical Chemistry group of the Institute for Molecules and Materials of the Radboud University who agreed to perform 
the computational calculations. At the moment of writing this proceeding, the Theoretical Chemistry group is finalizing the
calculations and the preliminary values of all the \meth ~molecule transitions, including all the masing transition, will soon
be available.

\end{document}